\documentclass[useAMS,usenatbib]{mn2e}

\usepackage{graphicx}
\usepackage{amssymb}
\usepackage{amsmath}
\usepackage{amstext}
\usepackage{multirow} 
\usepackage{lscape}

\newcommand{\elecd}{$n_{\rm e}$}
\newcommand{\elect}{$T_{\rm e}$}

\newcommand{\foiii}{[O\thinspace{\sc iii}]}

\newcommand{\foii}{[O\thinspace{\sc ii}]}
\newcommand{\fsii}{[S\thinspace{\sc ii}]}
\newcommand{\fsiii}{[S\thinspace{\sc iii}]}

\newcommand{\fnii}{[N\thinspace{\sc ii}]}
\newcommand{\fariv}{[Ar\thinspace{\sc iv}]}
\newcommand{\fcliii}{[Cl\thinspace{\sc iii}]}
\newcommand{\fclii}{[Cl\thinspace{\sc ii}]}
\newcommand{\fcliv}{[Cl\thinspace{\sc iv}]}

\newcommand{\fariii}{[Ar\thinspace{\sc iii}]}

\newcommand{\hii}{H\thinspace{\sc ii}}

\newcommand{\heii}{He\thinspace{\sc ii}}

\newcommand\ion[2]{${\rm #1^{#2}}$}           % ej. \ion{O}{+} --> O^+

\title[The Galactic abundance gradient of chlorine]{The radial abundance gradient of chlorine in the Milky Way}
\author[C. Esteban et al.]
       {C. Esteban$^{1,2}$\thanks{E-mail: cel@iac.es}, 
        J. Garc{\'{\i}}a-Rojas$^{1,2}$, V. P\'erez-Mesa$^{1,2}$\\
	$^1$Instituto de Astrof\'\i sica de Canarias, E-38200 La Laguna, Tenerife, Spain\\
        $^2$Departamento de Astrof\'\i sica, Universidad de La Laguna, E-38206, La Laguna, Tenerife, Spain\\}

\begin{document}

\date{Accepted 2015 June 17. Received 2015 June 5; in original form 2015 April 29}
\pagerange{\pageref{firstpage}--\pageref{lastpage}} \pubyear{2010}

\maketitle
\label{firstpage}

%%%%%%%%%%%%%%%%%%%%%%%%%%%%%%%%%%%%%%%%%%%%%%%%%%%%%%%%%%%%%%%%%%%%%%%%%%%%%%%%%%%%%%%%%%%%%%%%%
%%%%%%%%%%%%%%%%%%%%%%%%%%%%%%%%%%%%%%%%%%%%%%%%%%%%%%%%%%%%%%%%%%%%%%%%%%%%%%%%%%%%%%%%%%%%%%%%%
\begin{abstract}
 
We determine the radial abundance gradient of Cl in the Milky Way from {\hii} regions spectra. For the first time, the Cl/H ratios are computed by simply adding ionic abundances and not using an ionization correction factor ($ICF$). We use a collection of published very deep spectra of Galactic {\hii} regions. We have re-calculated the physical conditions, ionic and total abundances of Cl and O using the same methodology and updated atomic data for all the objects. We find that the slopes of the radial gradients of Cl and O are identical within the uncertainties: $-$0.043 dex kpc$^{-1}$. This is consistent with a lockstep evolution of both elements. We obtain that the mean value of the Cl/O ratio across the Galactic disc is log(Cl/O) = $-$3.42 $\pm$ 0.06.  We compare our Cl/H ratios with those determined from \ion{Cl}{2+} abundances and using some available $ICF$ schemes of the literature. We find that our total Cl abundances are always lower than the values determined using $ICF$s, indicating that those correction schemes systematically overestimate the contribution of \ion{Cl}{+} and \ion{Cl}{3+} species to the total Cl abundance. Finally, we propose an empirical $ICF$(\ion{Cl}{2+}) to estimate the Cl/H ratio in {\hii} regions. 

 \end{abstract}


%%%%%%%%%%%%%%%%%%%%%%%%%%%%%%%%%%%%%%%%%%%%%%%%%%%%%%%%%%%%%%%%%%%%%%%%%%%%%%%%%%%%%%%%%%%%%%%%%
\section{Introduction} \label{intro}
%%%%%%%%%%%%%%%%%%%%%%%%%%%%%%%%%%%%%%%%%%%%%%%%%%%%%%%%%%%%%%%%%%%%%%%%%%%%%%%%%%%%%%%%%%%%%%%%%

Chlorine (Cl) is located in Group VIIA of the periodic table. The elements of this group are known as halogens, which also includes fluorine, bromine and iodine. 
Cl is between the 17th and 20th most abundant element in the Solar System \citep{asplundetal09} and its most common isotopes are $^{35}$Cl and $^{37}$Cl. These two isotopes are 
produced during both hydrostatic and explosive oxygen burning in stars \citep{woosleyweaver95, pagel97}, although some contribution may be produced by Type Ia supernovae 
\citep{nomotoetal97}. \citet{clayton03} indicates that once $^{34}$S is formed during oxygen burning, a single additional proton capture yields $^{35}$Cl. In the case of $^{37}$Cl, most of it is produced 
by radioactive decay of $^{37}$Ar, which is formed by a single neutron capture by $^{36}$Ar. These processes seem to account for the 
solar $^{35}$Cl and $^{37}$Cl abundances \citep{clayton03}. 
Since lines of Cl are very rare in stellar spectra, there are very few determinations of its abundance in stars \citep{clayton03}. In addition, the Cl abundance is difficult to calculate in diffuse and dense interstellar clouds 
\citep{jenkins09}. Therefore, most of the information we have about Cl abundances comes from the analysis of emission lines in ionized nebulae. 

The determination of Cl abundances in {\hii} regions and planetary nebulae (PNe) is normally based on the measurement of the intensity of  
{\fcliii} doublet at 5517 and 5537 \AA, which are by far the brighest lines of Cl in optical spectra. For the common ionization conditions of {\hii} regions, we expect to have 
only significant amounts of \ion{Cl}{+}, \ion{Cl}{2+} and \ion{Cl}{3+}. Given the similarity between the ionization potential (IP) of \ion{He}{+} and \ion{Cl}{3+}, we do not expect a significant amount of \ion{Cl}{4+} in {\hii} regions with undetected {\heii} emission lines. The brightest lines of \ion{Cl}{+} and \ion{Cl}{3+} are {\fclii} 9123 \AA\ and {\fcliv} 8046 \AA, respectively  
and, for example, in the Orion Nebula are only between 4\% and 8\% the intensity of the {\fcliii} lines. Therefore, the detection of {\fclii} and {\fcliv} lines may only be possible 
in deep spectra of bright objects that also include the far red (until about 9100 \AA). Moreover, the reddest part of the optical spectrum is heavily contaminated by intense sky lines so 
the use of high spectral resolution or an exquisite sky emission removal is mandatory in order to separate these faint nebular lines. 
In the usual situation of only having {\fcliii} lines in the spectra we must adopt an ionization 
correction factor ($ICF$) to estimate the total Cl abundance. In this case, the $ICF$(\ion{Cl}{2+}) is a multiplicative factor to transform \ion{Cl}{2+}/\ion{H}{+} ratios into Cl/H ones. 
It depends on the ionization degree of the nebula, generally on its \ion{O}{+}/\ion{O}{2+} ratio or a combination of the \ion{O}{+}/\ion{O}{2+} and \ion{S}{+}/\ion{S}{2+} ones. 
For {\hii} region studies, there are different expressions for the $ICF$(\ion{Cl}{2+}) available in the literature. For example, \cite{peimberttorrespeimbert77} proposed a relation 
based on the similarity of the ionization potential of ionic species of Cl, O and S but others, as \cite{mathisrosa91} or \cite{izotovetal06}, obtained fitting expressions based on 
photoionization models. A very complete $ICF$ scheme for PNe has been recently presented by \cite{delgadoingladaetal14}. That work includes an $ICF$(\ion{Cl}{2+}/\ion{O}{+}) that can be applicable 
for {\hii} region conditions \citep{delgadoingladaetal15}. 

The radial abundance gradient of Cl in the Milky Way has only been studied in papers devoted to PNe but never in works based on {\hii} region observations. \cite{faundez-abansmaciel87}, 
\cite{macielchiappini94} and \cite{henryetal04} have determined the radial gradient of Cl in the Milky Way from abundance data for large samples of PNe, finding slopes from 
$-$0.07 to $-$0.04 dex kpc$^{-1}$. The last two works conclude that O and Cl gradients are similar, a result which is consistent with a lockstep evolution of both elements. 
However, some recent theoretical works indicate that PNe progenitors can modify their initial O abundance. Some O enrichment can be produced in the third dredge-up of asymptotic giant branch (AGB) stars 
\citep[e.g.][]{karakaslattanzio14} but, on the other hand, hot bottom burning in the most massive progenitors of PNe can destroy it  
\citep[e.g.][]{karakaslattanzio14}. The efficiency of both processes would depend on the mass of the progenitors stars and their metallicity. 
In any case, the observational evidences of such O variations in PNe are contradictory. \citet{delgadoingladaetal15} 
have obtained observational evidences of significant O enrichment (by $\sim$ 0.3 dex) in PNe with carbon-rich dust, with progenitor masses in the range 1.5$-$3 M$_{\odot}$. 
But, on the other hand, \cite{magrinietal09} comparing the O and Ne abundances of the PNe population of M33, did not find any evidence of systematic dredge-up of O in the metallicity range of that galaxy. Moreover, \cite{magrinigoncalves09} reviewed the O/H ratios of PNe and {\hii} regions in dwarf galaxies of the Local 
Group finding possible evidences of O dredge-up only at very low metallicities. In contrast, there is not any theoretical prediction that the amount of Cl may change along the evolution of the low and 
intermediate stars that produce PNe. In fact, \citet{delgadoingladaetal15} propose Cl/H ratio as a more reliable proxy for metallicity than O/H in PNe abundance studies. 
Therefore, it would be not unexpected to obtain somewhat different gradients for O and Cl using PNe abundance data. The aim of this paper is to determine the Cl and O abundance 
for a sample of {\hii} regions with the best available spectra and located across the disc of the Galaxy to better explore the nucleosynthetic origin of Cl. 

The structure of this paper is as follows. In \S\ref{data} we describe the set of {\hii} region spectra used to derive the nebular parameters. In \S\ref{results} we present 
the results of our physical conditions and abundances recalculations, as well discuss the results pertaining to the Galactic Cl radial gradients, the local Cl/O ratio and the 
empirical $ICF$(\ion{Cl}{2+}) we propose. Finally, in \S\ref{conclusions} we summarize our main conclusions.

%%%%%%%%%%%%%%%%%%%%%%%%%%%%%%%%%%%%%%%%%%%%%%%%%%%%%%%%%%%%%%%%%%%%%%%%%%%%%%%%%%%%%%%%%%%%%%%%%
\section{Observational data} \label{data}
%%%%%%%%%%%%%%%%%%%%%%%%%%%%%%%%%%%%%%%%%%%%%%%%%%%%%%%%%%%%%%%%%%%%%%%%%%%%%%%%%%%%%%%%%%%%%%%%%

We have used intensity ratios of several selected emission lines of spectra of Galactic {\hii} regions published by \citet{estebanetal04, estebanetal13} and  \citet{garciarojasetal04, garciarojasetal05, garciarojasetal06, garciarojasetal07}. The spectra were obtained with the Ultraviolet Visual Echelle Spectrograph 
\citep[UVES,][]{dodoricoetal00} 
at the Very Large Telescope (VLT). Nowadays, this dataset is the deepest and most complete collection of spectra of Galactic {\hii} regions available. The list of objects, their 
Galactocentric distances and the reference of their published spectra are presented in Table~\ref{sample}. All Galactocentric distances have been 
taken from \cite{estebanetal05} except in the case of NGC~2579. For this object we use the value adopted by \cite{estebanetal13} and calculated by \cite{copettietal07}. 
 The distances are based on kinematical or photometrical methods and have typical uncertainties of less than 1 kpc \citep[e.g.][]{russeil03}. We have assumed the Sun at 8 kpc from the Galactic Centre.

 %%%%%%%%%%%%%
 
  \begin{table}
   \begin{minipage}{170mm}
     \caption{Sample of Galactic {\hii} regions.}
     \label{sample}
    \begin{tabular}{lcc}
     \hline
     & $R_{\rm G}$ & \\     
     {\hii} region &  (kpc) &  Data source \\
     \hline
     M16 & 6.34 & \cite{garciarojasetal06} \\
     M8 & 6.41 & \cite{garciarojasetal07} \\
     M17 & 6.75 & \cite{garciarojasetal07} \\
     M20 & 7.19 & \cite{garciarojasetal06} \\
     NGC~3576 & 7.46 & \cite{garciarojasetal04} \\
     M42 & 8.40 & \cite{estebanetal04}\\
     NGC~3603 & 8.65 & \cite{garciarojasetal06} \\
     S~311 & 10.43 & \cite{garciarojasetal05} \\
     NGC~2579 & 12.40 & \cite{estebanetal13} \\
     \hline
    \end{tabular}
   \end{minipage}
  \end{table}

 %%%%%%%%%%%%%

%%%%%%%%%%%%%%%%%%%%%%%%%%%%%%%%%%%%%%%%%%%%%%%%%%%%%%%%%%%%%%%%%%%%%%%%%%%%%%%%%%%%%%%%%%%%%%%%%
\section{Results and discussion}\label{results}     
%%%%%%%%%%%%%%%%%%%%%%%%%%%%%%%%%%%%%%%%%%%%%%%%%%%%%%%%%%%%%%%%%%%%%%%%%%%%%%%%%%%%%%%%%%%%%%%%%

Although the source papers of the data used in this work present calculations of physical conditions -- electron temperature, {\elect}, and density, {\elecd} -- and ionic and total abundances 
of some elements, 
we prefer to re-calculate all the relevant quantities in order to have a homogeneous set of values calculated with the same methodology and updated atomic data. 

%%%%%%%%%%%%%

  \begin{table*}
  \centering
   \begin{minipage}{180mm}
     \caption{Atomic dataset used for collisionally excited lines.}
     \label{atomic}
    \begin{tabular}{lcc}
     \hline 
	& Transition probabilities &  \\
	Ion & and energy levels & Collisional strengths \\
     \hline 
N$^+$ & \cite{froesefischertachiev04} & \cite{tayal11} \\
O$^+$ & \cite{froesefischertachiev04} & \cite{kisieliusetal09} \\
O$^{2+}$ &  \cite{froesefischertachiev04, storeyzeippen00} & \cite{storeyetal14} \\
S$^+$ & \cite{Podobedovaetal09} & \cite{TayalZatsarinny10} \\
S$^{2+}$ &  \cite{Podobedovaetal09} & \cite{tayalgupta99} \\
Ar$^{2+}$ & \cite{mendoza83, kaufmansugar86} & \cite{galavisetal95} \\
Ar$^{3+}$ & \cite{mendozazeippen82b} & \cite{ramsbottombell97} \\
Cl$^{+}$ &  \cite{mendozazeippen83}  & \cite{tayal04} \\
Cl$^{2+}$ & \cite{mendoza83, kaufmansugar86} & \cite{butlerzeippen89} \\
& \cite{fritzscheetal99} & \cite{mendoza83}\\
Cl$^{3+}$ &  \cite{kaufmansugar86, ellismartinson84} & \cite{galavisetal95} \\
& \cite{mendozazeippen82a}&  \\
     \hline
    \end{tabular}
   \end{minipage}
  \end{table*}
 %%%%%%%%%%%%%
 
 \subsection{Physical Conditions } \label{conditions}
 
Physical conditions were calculated making use of the version 1.0.2 of {\sc PyNeb} \citep{Luridianaetal15}. {\sc PyNeb} is an updated and more versatile version 
of the {\sc nebular} package of {\sc iraf} written in python language. The density-sensitive emission line ratios we have used are {\foii}~3726/3729, {\fsii}~6717/6731, 
{\fcliii}~5518/5538 and {\fariv}~4711/4740, and the temperature-sensitive ones are {\fnii}~5755/(6548+6584), {\foii}~(7319+7330)/(3726+3729), 
{\fsii}~(4069+4076)/(6717+6731), {\foiii}~4363/(4959+5007), {\fsiii}~6312/(9069+9532), and {\fariii}~5192/(7136+7751). {\sc PyNeb}  permits to easily customize the atomic datasets. 
In our case we have used the atomic data listed in Table~\ref{atomic}. Temperatures and densities were determined using an iterative procedure. 
Firstly, we assume an initial {\elect} of 10,000 K to determine {\elecd} and this last value is used to calculate {\elect}. The uncertainties of each quantity 
were obtained performing Monte Carlo calculations changing the values of the emission line ratios within the line intensity errors quoted in the source papers of 
the observational data (see \S\ref{abundances}). Temperatures of low ionization 
species -- {\elect}({\foii}), {\elect}({\fnii}) and {\elect}({\fsii}) -- were derived considering the average of {\elecd}({\foii}) and {\elecd}({\fsii}). For calculating {\elect}({\foiii}), 
{\elect}({\fsiii}) and {\elect}({\fariii}) we considered the average of all the density determinations available -- {\elecd}({\foii}), {\elecd}({\fsii}), {\elecd}({\fcliii}) and 
{\elecd}({\fariv}) --
because all the density-sensitive ratios give similar values and also because {\elecd}({\foii}) and {\elecd}({\fsii}) show the lowest uncertainties. Using an average of 
{\elecd}({\fcliii}) and {\elecd}({\fariv}) would provide almost identical results but with larger uncertainties.  
Tables~\ref{temperatures} and ~\ref{densities} show the values of electron temperatures and densities, respectively, obtained from the different 
individual line-intensity ratios. Considering that \ion{Cl}{2+} is the the dominant ionic species of Cl in our {\hii} region sample and that there are several sets of atomic data available, 
we decided to use combinations of those datasets for deriving the {\elecd}({\fcliii}) as well as the \ion{Cl}{2+}/\ion{H}{+} ratio. In Table~\ref{sets} we present the different combinations used for 
this ion. The value of {\elecd}({\fcliii})  included in Table~\ref{densities} is the average of the densities obtained with the six combinations. The final adopted electron density for each object, 
$<${\elecd}$>$, is the weighted mean of the values obtained for the different density-sensitive indicators and it is also included in Table~\ref{densities}.

 %%%%%%%%%%%%%
 \begin{table}
  \centering
   \begin{minipage}{180mm}
     \caption{Combinations of atomic datasets used for \ion{Cl}{2+}.}
     \label{sets}
    \begin{tabular}{lcc}
     \hline 
	& Transition probabilities &  \\
	Set No. & and energy levels & Collisional strengths \\
     \hline 
1 & \cite{mendoza83} & \cite{butlerzeippen89} \\
2 & \cite{mendoza83} & \cite{mendoza83} \\
3 & \cite{mendoza83} & \cite{butlerzeippen89} \\
& \cite{kaufmansugar86} & \\
4 & \cite{mendoza83} & \cite{mendoza83} \\
& \cite{kaufmansugar86} & \\
5 & \cite{fritzscheetal99} & \cite{butlerzeippen89} \\
6 & \cite{fritzscheetal99} & \cite{mendoza83} \\
     \hline
    \end{tabular}
   \end{minipage}
  \end{table}
 %%%%%%%%%%%%%

We assume a two-zone approximation for the nebula estimating the representative values of the electron temperature for the zones where low and high-ionization potential ions 
are present, {\elect}(low) and {\elect}(high). Those values have been used to determine ionic abundances. {\elect}(low) was calculated as the mean of {\elect}({\fnii}), {\elect}({\foii}) 
and {\elect}({\fsii}) weighted by their uncertainties. On the other hand, {\elect}(high) was calculated as the weighted mean of 
{\elect}({\fsiii}), {\elect}({\fariii}) and {\elect}({\foiii}). The corresponding values of {\elect}(low)  and {\elect}(high) used for each object are included in 
Table~\ref{temperatures}.

%%%%%%%%%%%%%
  \begin{table*}
   \centering
   \begin{minipage}{180mm}
   \caption{Electron temperatures (K).}
   \label{temperatures}
    \begin{tabular}{lcccccccc}
     \hline
     Object & {\elect}({\fnii}) & {\elect}({\foii}) & {\elect}({\fsii}) & {\elect}(low) & {\elect}({\foiii}) & {\elect}({\fsiii}) &
     {\elect}({\fariii}) & {\elect}(high) \\  
     \hline
	M16 & 8360 $\pm$ 150 & 8330 $\pm$ 1210 & 9290 $\pm$ 730 & {\bf 8400 $\pm$ 190} & 7640 $\pm$ 140 & 8300 $\pm$ 270 & $-$ & {\bf 7780 $\pm$ 170} \\ 
	M8 & 8460 $\pm$ 130 & 9370 $\pm$ 2410 & 9600 $\pm$ 740 & {\bf 8520 $\pm$ 210} & 8060 $\pm$ 100 & 8460 $\pm$ 170 & 7520 $\pm$ 310 & {\bf 8120 $\pm$ 130} \\ 
	M17 & 8890 $\pm$ 250 & 9000 $\pm$ 5130 & 8290 $\pm$ 650 & {\bf 8810 $\pm$ 310} & 8000 $\pm$ 120 & 8070 $\pm$ 230 & 8380 $\pm$ 410 & {\bf 8040 $\pm$ 160} \\ 
	M20 & 8340 $\pm$ 160 & 8560 $\pm$ 1670 & 9050 $\pm$ 500 & {\bf 8410 $\pm$ 200} & 7870 $\pm$ 260 & 8080 $\pm$ 210& $\pm$ & {\bf 7990 $\pm$ 250} \\ 
	NGC~3576 & 8680 $\pm$ 210 & 10090 $\pm$ 1850 & 11280 $\pm$ 1620 & {\bf 8740 $\pm$ 250} & 8460 $\pm$ 130 & 8650 $\pm$ 400 & 8570 $\pm$ 360 & {\bf 8490 $\pm$ 180} \\ 
	M42 & 10290 $\pm$ 300 & 10530 $\pm$ 1770 & 13610 $\pm$ 3550 & {\bf 10320 $\pm$ 360} & 8350 $\pm$ 90 & 9520 $\pm$ 580 & 8290 $\pm$ 260& {\bf 8370 $\pm$ 120} \\ 
	NGC~3603 & 11360 $\pm$ 600 & 18980 $\pm$ 4370 & 18870 $\pm$ 3090 & {\bf 11760 $\pm$ 760} & 9080 $\pm$ 150 & 9670 $\pm$ 360 & $-$ & {\bf 9170 $\pm$ 180} \\ 
	S~311 & 9370 $\pm$ 210 & 10330 $\pm$ 1950 & 9610 $\pm$ 500 & {\bf 9420 $\pm$ 270} & 8980 $\pm$ 110 & 9470 $\pm$ 280 & 8760 $\pm$ 970 & {\bf 9040 $\pm$ 140} \\ 
	NGC~2579 & $-$ & 12460 $\pm$ 1690 & 12240 $\pm$ 1230 & {\bf 12340 $\pm$ 1280} & 9370 $\pm$ 60 & 10400 $\pm$ 290 & 8640 $\pm$ 290 & {\bf 9380 $\pm$ 80} \\ 
     \hline
    \end{tabular} 
   \end{minipage}
  \end{table*}
  %%%%%%%%%%%%%      

%%%%%%%%%%%%%
  \begin{table*}
   \centering
   \begin{minipage}{180mm}
   \caption{Electron densities (cm$^{-3}$).}
   \label{densities}
    \begin{tabular}{lccccc}
     \hline
     Object & {\elecd}({\fsii}) & {\elecd}({\foii}) & {\elecd}({\fcliii}) & {\elecd}({\fariv}) & $<${\elecd}$>$ \\
     \hline
	M16 & 1100 $\pm$ 200 & 1140 $\pm$ 230 & 1215 $\pm$ 560 & $-$ & {\bf 1120 $\pm$ 240} \\ 
	M8 & 1280 $\pm$ 160 & 1620 $\pm$ 630 & 1930 $\pm$ 320 & 1990: & {\bf 1420 $\pm$ 210} \\ 
	M17 & 780 $\pm$ 180 & 500 $\pm$ 110 & 250 $\pm$ 240 & $-$ & {\bf 530 $\pm$ 140} \\
	M20 & 270 $\pm$ 80 & 260 $\pm$ 50 & 320 $\pm$ 290 & $-$ & {\bf 260 $\pm$ 60} \\
	NGC~3576 & 1090 $\pm$ 320 & 1700 $\pm$ 380 & 3180 $\pm$ 710 & 2930 $\pm$ 1450 & {\bf 1570 $\pm$ 410} \\
	M42 & 4590 $\pm$ 3910 & 6590 $\pm$ 1970 & 7470 $\pm$ 750 & 4870 $\pm$ 1090 & {\bf 6590 $\pm$ 1020} \\ 
	NGC~3603 & 3000 $\pm$ 1260 & 2600 $\pm$ 650 & 4850 $\pm$ 1710 & 928: & {\bf 2910 $\pm$ 870} \\ 
	S~311 & 300 $\pm$ 90 & 280 $\pm$ 80 & 500 $\pm$ 420 & $-$ & {\bf 290 $\pm$ 90} \\ 
	NGC~2579 & 920 $\pm$ 170 & 1240 $\pm$ 280 & 1790 $\pm$ 550 & 1030: & {\bf 1060 $\pm$ 220} \\ 
     \hline
    \end{tabular} 
   \end{minipage}
  \end{table*}
  %%%%%%%%%%%%%       

\subsection{Ionic and total abundances} \label{abundances}

Ionic abundances of \ion{O}{+}, \ion{O}{2+}, \ion{Cl}{+},  \ion{Cl}{2+} and \ion{Cl}{3+} have been derived from CELs under the two-zone scheme using the {\sc PyNeb} package  
and the atomic dataset indicated in Table~\ref{atomic}. We have used {\elect}(low) for calculating  \ion{O}{+} and  \ion{Cl}{+} and {\elect}(high) for  \ion{O}{2+},  \ion{Cl}{2+} and \ion{Cl}{3+}. Although there are several datasets of atomic data for \ion{O}{+} and \ion{O}{2+} available in the literature, we have selected those that, according to our experience, provide more reliable results.
In the case of \ion{Cl}{2+}, considering that a) it is the dominant ionic species of Cl in our {\hii} region sample, and b) the availability of several references of atomic data, 
we decided to calculate the \ion{Cl}{2+}/\ion{H}{+} ratio using different combinations of datasets, which are included in Table~\ref{sets} and indicated with different numbers. 
In contrast, for  \ion{Cl}{+} and \ion{Cl}{3+} we have a single option of atomic dataset available and therefore no other combinations are possible.

In Table~\ref{abund_Cl++} we present the \ion{Cl}{2+}/\ion{H}{+} obtained for the different combinations of atomic data used 
and the weighted mean value that we assume as representative for the abundance of this ion for each {\hii} region. In Table~\ref{abund} we include all the ionic abundances we calculate for the different 
ionic species of Cl and O which emission lines have been detected in the nebulae. The uncertainties of ionic abundances were obtained performing Monte Carlo calculations. To do that, we generate 500 random values for each line intensity using a Gaussian distribution centered in the observed line intensity with a sigma equal to the associated uncertainty. For higher number of Monte Carlo simulations, the errors in the computed quantities remain constant.

The total abundances of Cl and O have been determined by simply adding their ionic abundances and are included in the last two columns of Table~\ref{abund}. The four nebulae of lower ionization degree of the sample do not show detections of {\fcliv} lines and we do not expect any relevant contribution of \ion{Cl}{3+} to the total Cl abundance. In Table~\ref{contCl3+} we compare the total Cl abundance we obtain from the sum of 
\ion{Cl}{+},  \ion{Cl}{2+} and \ion{Cl}{3+} abundances with that we obtain only adding \ion{Cl}{+} and \ion{Cl}{2+} and also the ionization degree of the nebulae (parametrized by the \ion{O}{2+}/\ion{O}{+} ratio). As we can 
see, the contribution of the \ion{Cl}{3+}/\ion{H}{+} ratio to the total Cl abundance is always very small even in the objects of the highest ionization degree and never larger than 0.02 dex (see column 3 of Table~\ref{contCl3+}). 
Therefore, the contribution of \ion{Cl}{3+} in the four {\hii} regions with the lowest \ion{O}{2+}/\ion{O}{+} ratios: M20, M16, M8 and S~311, should be negligible. For these four objects, the assumption of 
Cl/H = \ion{Cl}{+}/\ion{H}{+}+\ion{Cl}{2+}/\ion{H}{+} seems to be perfectly applicable. 

   %%%%%%%%%%%%%
  \begin{table*}
   \centering
   \begin{minipage}{180mm}
     \caption{ \ion{Cl}{2+}/ \ion{H}{+} ratios$^{\rm a}$ determined from combinations of atomic data sets and adopted mean value}
     \label{abund_Cl++}
    \begin{tabular}{lcccccccc}
     \hline
        Object & 1 & 2 & 3 & 4 & 5 & 6 & Adopted & \\
     \hline
	M16 & 5.06 $\pm$ 0.04 &  5.08 $\pm$ 0.04 &  5.05 $\pm$ 0.04 &  5.08 $\pm$ 0.04 &  5.06 $\pm$ 0.04 &  5.08 $\pm$ 0.04 & {\bf 5.07 $\pm$ 0.04} \\ 
	M8 & 5.03 $\pm$ 0.02 &  5.05 $\pm$ 0.02 &  5.02 $\pm$ 0.02 &  5.04 $\pm$ 0.04 &  5.02 $\pm$ 0.02 &  5.05 $\pm$ 0.03 &  {\bf 5.04 $\pm$ 0.03} \\
	M17 & 5.05 $\pm$ 0.03 &  5.07 $\pm$ 0.03 &  5.05 $\pm$ 0.03 &  5.07 $\pm$ 0.03 &  5.05 $\pm$ 0.03 &  5.07 $\pm$ 0.03 &  {\bf 5.06 $\pm$ 0.03} \\
	M20 & 4.99 $\pm$ 0.04 &  5.01 $\pm$ 0.04 &  4.99 $\pm$ 0.04 &  5.01 $\pm$ 0.04 &  4.99 $\pm$ 0.04 &  5.02 $\pm$ 0.04 &   {\bf 5.00 $\pm$ 0.04} \\
	NGC~3576 & 4.93 $\pm$ 0.03 &  4.96 $\pm$ 0.04 &  4.93 $\pm$ 0.04 &  4.95 $\pm$ 0.05 &  4.93 $\pm$ 0.04 &  4.96 $\pm$ 0.05 &  {\bf 4.94 $\pm$ 0.04} \\
	M42 & 5.12 $\pm$ 0.03 &  5.14 $\pm$ 0.02 &  5.11 $\pm$ 0.03 &  5.13 $\pm$ 0.05 &  5.11 $\pm$ 0.02 &  5.13 $\pm$ 0.03 &  {\bf 5.12 $\pm$ 0.03} \\ 
	NGC~3603 & 5.02 $\pm$ 0.05 &  5.05 $\pm$ 0.05 &  5.02 $\pm$ 0.06 &  5.04 $\pm$ 0.06 &  5.02 $\pm$ 0.05 &  5.04 $\pm$ 0.06 & {\bf 5.03 $\pm$ 0.05} \\  
	S~311 & 4.85 $\pm$ 0.03 &  4.88 $\pm$ 0.03 &  4.85 $\pm$ 0.03 &  4.87 $\pm$ 0.03 &  4.85 $\pm$ 0.04 &  4.88 $\pm$ 0.03 &  {\bf 4.86 $\pm$ 0.03} \\ 
	NGC~2579 & 4.80 $\pm$ 0.03 &  4.83 $\pm$ 0.03 &  4.80 $\pm$ 0.03 &  4.82 $\pm$ 0.04 &  4.80 $\pm$ 0.03 &  4.83 $\pm$ 0.03 &  {\bf 4.81 $\pm$ 0.03} \\
     \hline
    \end{tabular} 
     \begin{description}
      \item[$^{\rm a}$] In units of 12+log(\ion{Cl}{2+}/\ion{H}{+}).   
    \end{description}      \end{minipage}
  \end{table*}
  %%%%%%%%%%%%%   
  
  %%%%%%%%%%%%%
  \begin{table*}
   \centering
   \begin{minipage}{180mm}
     \caption{Ionic and total abundances for O and Cl$^{\rm a}$}
     \label{abund}
    \begin{tabular}{lcccccccc}
     \hline
        Object & \ion{O}{+} & \ion{O}{2+} & \ion{Cl}{+} & \ion{Cl}{2+} & \ion{Cl}{3+} & O &  Cl \\
     \hline
	M16 & 8.44 $\pm$ 0.04 &  7.88 $\pm$ 0.03 &  4.42 $\pm$ 0.04 &  5.07 $\pm$ 0.04 &  $-$ &  8.54 $\pm$ 0.04 &  {\bf 5.15 $\pm$ 0.04} \\ 
	M8 & 8.32 $\pm$ 0.06 &  7.87 $\pm$ 0.02 &  3.99 $\pm$ 0.23 &  5.04 $\pm$ 0.03 &  $-$ &  8.45 $\pm$ 0.04 &  {\bf 5.09 $\pm$ 0.03} \\ 
	M17 & 7.81 $\pm$ 0.05 &  8.43 $\pm$ 0.03 &  3.59 $\pm$ 0.10 &  5.06 $\pm$ 0.03 &  3.10 $\pm$ 0.22 &  8.53 $\pm$ 0.02 &  {\bf 5.08 $\pm$ 0.03} \\ 
	M20 & 8.44 $\pm$ 0.04 &  7.68 $\pm$ 0.04 &  4.44 $\pm$ 0.04 &  5.00 $\pm$ 0.04 &  $-$ &  8.51 $\pm$ 0.04 &  {\bf 5.11 $\pm$ 0.04} \\
	NGC~3576 & 8.09 $\pm$ 0.06 &  8.36 $\pm$ 0.03 &  3.80 $\pm$ 0.06 &  4.94 $\pm$ 0.04 &  3.21 $\pm$ 0.05 &  8.55 $\pm$ 0.03 &  {\bf 4.98 $\pm$ 0.04} \\
	M42 & 7.75 $\pm$ 0.06 &  8.42 $\pm$ 0.02 &  3.49 $\pm$ 0.07 &  5.12 $\pm$ 0.03 &  3.65 $\pm$ 0.05 &  8.50 $\pm$ 0.02 &  {\bf 5.15 $\pm$ 0.03} \\ 
	NGC~3603 & 7.29 $\pm$ 0.09 &  8.41 $\pm$ 0.03 &  3.13 $\pm$ 0.06 &  5.03 $\pm$ 0.05 &  3.85 $\pm$ 0.03 &  8.44 $\pm$ 0.03 &  {\bf 5.06 $\pm$ 0.05} \\  
	S~311 & 8.25 $\pm$ 0.05 &  7.81 $\pm$ 0.02 &  4.29 $\pm$ 0.04 &  4.86 $\pm$ 0.03 &  $-$ &  8.39 $\pm$ 0.03 &  {\bf 4.96 $\pm$ 0.03} \\ 
	NGC~2579 & 7.35 $\pm$ 0.12 &  8.20 $\pm$ 0.02 &  3.25 $\pm$ 0.07 &  4.81 $\pm$ 0.03 &  2.78 $\pm$ 0.05 &  8.26 $\pm$ 0.03 &  {\bf 4.83 $\pm$ 0.03} \\ 
     \hline
    \end{tabular} 
     \begin{description}
      \item[$^{\rm a}$] In units of 12+log(\ion{X}{i+}/\ion{H}{+}) or 12+log(X/H).   
    \end{description}   
  \end{minipage}
  \end{table*}
  %%%%%%%%%%%%%   

  \subsection{The Galactic Cl/H gradient} \label{clgradient}

We have performed least-squares linear fits to the galactocentric distance of the {\hii} regions, $R_{\rm G}$, and their corresponding Cl/H, O/H and Cl/O ratios. 
For the fitting, each elemental ratio has been weighted by the inverse of its uncertainty. These fits give the following radial gradients: 
 \begin{equation}
12 + \log(\mathrm{Cl/H}) = 5.40(\pm 0.11) - 0.043(\pm 0.012) \times R_{\mathrm G};
\end{equation}
\begin{equation}
12 + \log(\mathrm{O/H}) = 8.83(\pm 0.07) - 0.043(\pm 0.009) \times R_{\mathrm G};
\end{equation}
\begin{equation}
\log(\mathrm{Cl/O}) = -3.41(\pm 0.12) - 0.001(\pm 0.014) \times R_{\mathrm G}.
\end{equation}

%%%%%%%%%%%%%
  \begin{table}
   \centering
   \begin{minipage}{180mm}
   \caption{Contribution of \ion{Cl}{3+}/\ion{H}{+} to the Cl/H ratio.}
   \label{contCl3+}
    \begin{tabular}{lcccc}
     \hline
     Object & Cl$^{\rm a}$ & \ion{Cl}{+}+\ion{Cl}{2+}$^{\rm b}$ & dif.$^{\rm c}$ & log(\ion{O}{2+}/\ion{O}{+}) \\
     \hline
	M20 & 5.11 & 5.11 & $-$ & $-$0.76 \\ 
	M16 & 5.15 & 5.15 & $-$ & $-$0.56 \\ 
	M8 & 5.09 & 5.09 & $-$ & $-$0.45 \\
	S~311 & 4.96 & 4.96 & $-$ & $-$0.44 \\
	NGC~3576 & 4.98 & 4.97 & 0.01 & 0.27 \\
	M17 & 5.08 & 5.07 & 0.01 & 0.62 \\ 
	M42 &  5.15 & 5.13 & 0.02 & 0.67 \\ 
	NGC~2579 & 4.83 & 4.82 & 0.01 & 0.85 \\ 
	NGC~3603 & 5.06 & 5.04 & 0.02 & 1.12 \\ 
     \hline
    \end{tabular} 
    \begin{description}
      \item[$^{\rm a}$] In units of 12+log(Cl/H).
      \item[$^{\rm b}$] In units of 12+log(\ion{Cl}{+}/\ion{H}{+}+\ion{Cl}{2+}/\ion{H}{+}).
      \item[$^{\rm c}$] Difference between columns 2 and 3.    
    \end{description}   
    \end{minipage}
  \end{table}
  %%%%%%%%%%%%%       

   %%%%%%%%%%%%%       
\begin{table} 
 \begin{minipage}{140mm}
 \caption{Determinations of the Galactic Cl/H abundance gradient.}
 \label{gradients}
  \begin{tabular}{l c c c}
\hline
 & & Slope & \\
 Reference & Objects & (dex kpc$^{-1}$) & Intercept \\
\hline
F-AM87$^{\rm a}$  & PNe & $-$0.057 & 5.67  \\ 
MC94$^{\rm b}$  & PNe & $-$0.07 $\pm$ 0.01 & $-$ \\
HKB04$^{\rm c}$  & PNe & $-$0.045 $\pm$ 0.013 & 5.45 $\pm$ 0.11  \\
This work & {\hii} regions & $-$0.043 $\pm$ 0.012 & 5.40 $\pm$ 0.11  \\
\hline
 \end{tabular}
    \begin{description}
      \item[$^{\rm a}$ ] \cite{faundez-abansmaciel87}.
      \item[$^{\rm b}$ ] \cite{macielchiappini94}.
      \item[$^{\rm c}$ ] \cite{henryetal04}.    
    \end{description}   
     \end{minipage}
\end{table} 
  %%%%%%%%%%%%%       

%%%%%%%%%%%%%%%
  \begin{figure}
   \centering
   \includegraphics[scale=0.5]{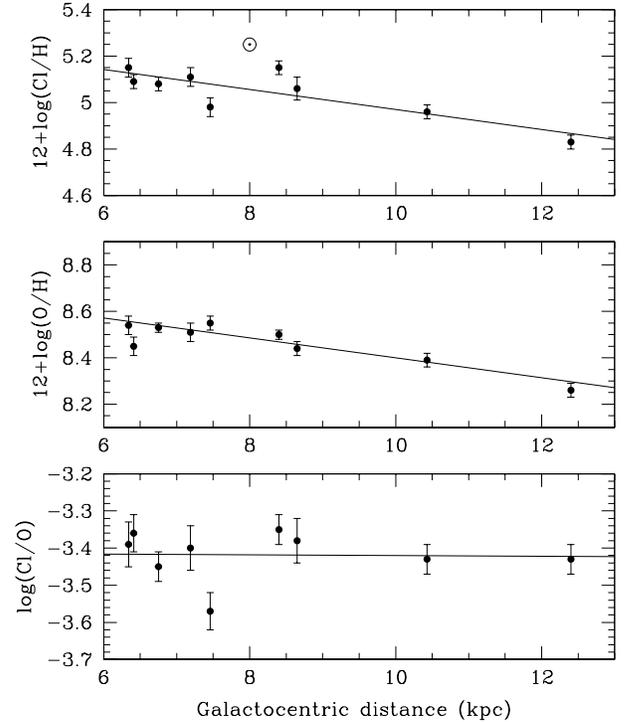} 
   \caption{Ionised gas phase Cl, O and Cl/O abundances of the Galactic {\hii} regions 
   studied in this work. The lines indicate the radial gradients obtained  
   from the linear fit of the data. The solar symbol indicates the meteoritic Cl abundance given by \citet{loddersetal09}. The Sun is located at 8 kpc.}
   \label{fig_gradients}
  \end{figure}    	 
 %%%%%%%%%%%%%%%

In Figure~\ref{fig_gradients} we represent the ionized gas phase Cl/H, O/H and Cl/O ratios of the {\hii} regions and the corresponding radial abundance gradients given above. 
In Table~\ref{gradients} we compare our Cl gradient with previous ones determined from PNe data. The slope of our Cl/H gradient is somewhat flatter than those determined by \citet{faundez-abansmaciel87} and \citet{macielchiappini94}. This may indicate an evolutionary effect, because a temporal flatting of the radial abundance gradients 
has been predicted in different chemical evolution models \citep[e.g.][]{houetal00, molladiaz05}. {\hii} regions abundances represent the present-day composition of the ISM but, in PNe, abundances of most elements should reflect the ISM composition in the past, at the moment of the formation of their progenitor stars, which have a wide age span, from about 1 Gyr to about 8 Gyr \citep[e.g.][]{macieletal09}. However, this possible temporal flattening of the Cl gradient is doubtful because the 
parameters of our fit shown in Figure~\ref{fig_gradients} -- slope and intercept -- are in very good agreement with those obtained by \citet{henryetal04}.

As we can see comparing the slopes of the gradients we obtain, the Cl/H and O/H gradients are identical within the uncertainties and this is reflected in a constant Cl/O ratio across the galactic disc. This is consistent with a lockstep evolution of both elements. As it was commented in \S\ref{intro}, a similar conclusion was obtained by \citet{macielchiappini94} and \citet{henryetal04} from large datasets of 
PNe spectra. \citet{clayton03} and \citet{henryetal04} comment that theoretically a certain amount of Cl may be produced by Type I supernova, but our results indicate that this amount should be negligible compared to the contribution due to massive stars. Theoretical predictions also indicate that O should also be in part produced by Type I supernova, but their relative contribution is only of the order of 8\% \citep[e.g.][]{woosleyetal02}.

In Figure~\ref{fig_gradients} we can see that two objects show somewhat discrepant Cl/H and Cl/O ratios with respect to the others. One is NGC~3576, at $R_{\rm G}$ = 7.46 kpc, and the other M42, the Orion nebula, at $R_{\rm G}$ = 8.40 kpc. 
In the case of NGC~3576, the rather low Cl/O ratio is produced by the combination of a low Cl/H and a somewhat high value of O/H. In the case of M42 the reason seems to be 
basically a higher value of Cl/H. It is difficult to ascertain the reason of such differences but they might be due to relative different contribution of local sources of O and Cl in the chemical composition of the nebulae or a different relative depletion fraction onto dust particles in those two objects.

\subsection{Comparison with the solar Cl abundance and the local Cl/O ratio} \label{comparison}

\citet{asplundetal09} indicate that the solar photospheric Cl abundance is very uncertain, 12+log(Cl/H) = 5.50 $\pm$ 0.30. This value cannot be derived from spectroscopy of the quiet Sun but, instead it was determined from IR spectrum of sunspots using 
lines of HCl \citep{hallnoyes72}. Meteoritic values presented by \citet{loddersetal09} are more confident, they give a value of 5.25 $\pm$ 0.06, which is included in Figure~\ref{fig_gradients} for comparison. From our radial Cl/H gradient, we estimate a 
ionized gas phase 12 + log(Cl/H) = 5.06 at the solar Galactocentric distance of 8 kpc. Therefore, the meteoritic Cl abundance is of the order of 0.2 dex higher than the nebular one. This is an apparently puzzling situation because meteorites should represent the composition of the solar nebula $\sim$4.5 Gyr ago and one would expect that the Cl abundance at the present-day ISM to be higher than the meteoritic value, just the opposite of what we find. Since -- as far as we know -- there are not numerical determinations about the evolution of Cl in the disc of the Milky Way, one would expect some increase of the Cl/H ratio in the last $\sim$5 Gyr. In fact, chemical evolution models by \cite{carigi03} and \cite{akermanetal04} estimate that the O/H ratio at the solar Galactocentric distance has increased by 0.12 dex since the Sun was formed. It is important to remark that the same problem exists when one compares the solar O/H ratio with that of ionized nebulae in the solar neighborhood. The solar photospheric O/H ratio is 
about 0.2 dex higher than the nebular values when determined from 
the intensity of collisionally excited lines (CELs) of ionic species of O. The reason of such difference can be due to depletion onto dust grains in the {\hii} regions and/or physical mechanisms associated to the so-called abundance discrepancy problem \citep[e.g.][]{garciarojasesteban07}. According to \citet{mesadelgadoetal09} and 
\citet{apeimbertpeimbert10}, the fraction of O embedded in dust grains in the Orion nebula is about 0.12 dex and, therefore, can only account for part of the problem. On the other hand, the abundance trend of Cl in diffuse and dense interstellar clouds is rather irregular and amount of dust depletion can not be estimated   
for this element \citep{jenkins09}. The abundance discrepancy problem arises when ones compares abundances of a given ion obtained from CELs and recombination lines (RLs). In all the {\hii} regions included in this paper, \ion{O}{2+} abundances determined from CELs are between 0.2 and 0.3 dex higher than those determined from RLs \citep[see][]{garciarojasesteban07}. Among different possibilities, one of the mechanisms that may produce such discrepancy is the presence of fluctuations in the 
spatial distribution of electron temperature in the nebulae \citep[the so-called temperature fluctuations,][]{peimbert67}. Under the presence of temperature fluctuations, ionic abundances determined from CELs would be lower than the real ones. 
A qualitatively similar effect would produce the presence of a ``kappa-distribution'' in the energy spectrum of free electrons in {\hii} regions, another explanation for the abundance discrepancy problem \citep{nichollsetal12}. In the presence of 
any of both mechanisms, temperature fluctuations or a ``kappa-distribution", the effect on the derived Cl/H would be similar, the Cl/H ratio we have determined in this paper -- from the intensity of CELs -- would be lower than the real ones, and therefore 
it can explain the difference between solar and nebular Cl abundances. 

From our data, we have determined the weighted mean of the Cl/O ratio across the Galactic disc and this is log(Cl/O) = $-$3.42 $\pm$ 0.06. \citet{henryetal04} compile a number of determinations of the Cl/O ratio based on PNe and 
{\hii} region observations, and the logarithmic values range from $-$3.67 to $-$3.30. Therefore, our mean ratio is consistent with those previous calculations. Moreover, from the list of recommended present-day solar abundances 
from photospheric and meteoritic data compiled by \citet{loddersetal09}, we obtain a log(Cl/O)$_{\odot}$ = $-$3.48 $\pm$ 0.09. In rather good agreement with our nebular determination. This similarity and the rather small dispersion of the individual values 
of the Cl/O ratio we obtain for the different {\hii} regions suggest that the individual or combined action of the two aforementioned effects -- dust depletion and abundance discrepancy -- should affect similarly to both elements -- Cl and O -- in all objects. 
  
  \subsection{An empirical ionization factor for Cl} \label{icfcl}

%%%%%%%%%%%%%
  \begin{table*}
   \centering
   \begin{minipage}{180mm}
   \caption{Comparion of Cl abundances determined with different $ICF$ schemes.}
   \label{compicfcl}
    \begin{tabular}{l c c c c c c}
     \hline
     & & \multicolumn{5}{c}{12 + $\log$(Cl/H)} \\
     Object & O/O$^{2+}$ & This work & P\&T-P77$^{\rm a}$ &  M\&R91$^{\rm b}$ & ISMGT06$^{\rm c}$ & D-IMS14$^{\rm d}$ \\
     \hline
	M20 & 6.76 & 5.11 & 5.18 & 5.12 & 5.18 & 5.17  \\ 
	M16 & 4.57 & 5.15 & 5.33& 5.24 & 5.23 & 5.22 \\ 
	M8 & 3.80 & 5.09 & 5.23& 5.13 & 5.18 & 5.18 \\ 
	S~311 & 3.80 & 4.96 & 5.12& 5.01 & 5.00 & 5.00 \\ 
	NGC~3576 & 1.55 & 4.98 & $-$ & 5.18 & 4.99 & 5.06 \\ 
	M17 & 1.26 & 5.08 & $-$ & 5.42 & 5.14 & 5.22 \\ 
	M42 &  1.20 & 5.15 & $-$ & 5.54 & 5.22 & 5.29 \\ 
	NGC~2579 & 1.15 & 4.83 & $-$ & 5.52 & 4.98 & 5.01 \\ 
	NGC~3603 & 1.07 & 5.06 & $-$ & 5.52 & 5.36 & 5.28 \\  
     \hline
    \end{tabular} 
    \begin{description}
      \item[$^{\rm a}$] \citet{peimberttorrespeimbert77}.
      \item[$^{\rm b}$] \citet{mathisrosa91}.
      \item[$^{\rm c}$] \citet{izotovetal06}. 
      \item[$^{\rm d}$] \citet{delgadoingladaetal14}
    \end{description}   
    \end{minipage}
  \end{table*}
  %%%%%%%%%%%%%       

%%%%%%%%%%%%%%%
  \begin{figure}
   \centering
   \includegraphics[scale=0.45]{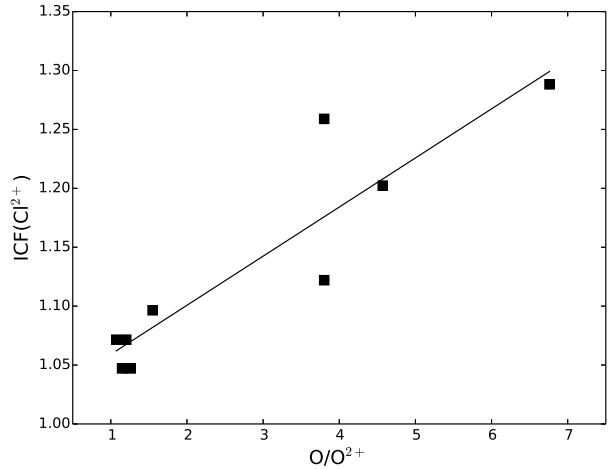} 
   \caption{Values of $ICF$(\ion{Cl}{2+}) as a function of the O/\ion{O}{2+} ratio for our {\hii} regions sample. The continuous line represents a linear fit to the data. }
   \label{fig_icf}
  \end{figure}    	 
 %%%%%%%%%%%%%%%

Usually, {\fcliii} lines are the only emission lines of Cl that can be measured in the spectra of {\hii} regions. Therefore, an ionization correction factor ($ICF$) has to be adopted to estimate the 
total Cl abundance. The $ICF$(\ion{Cl}{2+}) is a multiplicative factor to transform \ion{Cl}{2+}/\ion{H}{+} ratios into Cl/H ones following the simple relation:
 \begin{equation}
\mathrm{Cl/H} = ICF(\mathrm{Cl}^{2+}) \times \mathrm{Cl}^{2+}/\mathrm{H}^+.
\end{equation}
 \citet{peimberttorrespeimbert77} proposed a relation 
based on the similarity of the ionization potential of the ionic species of Cl with those of O and S. However, they only recommend its use for values of $ICF$(\ion{Cl}{2+}) close to one. 
\citet{mathisrosa91} presented a quite complete $ICF$ scheme for several elements based on nebular photoionization models using mainly plane-parallel stellar 
atmospheres as ionization sources. These authors admit that their $ICF$(\ion{Cl}{2+}) were very uncertain, with abundance differences of factor of two for a given object 
depending on the model atmosphere used. \citet{izotovetal06} obtained fitting expressions based on photoionization models for extragalactic {\hii} regions 
and defined different fittings depending on three metallicity ranges. Finally, \cite{delgadoingladaetal14} 
compute a large grid of photoionization models covering a wide range of physical parameters representative of PNe. This work includes an $ICF$(\ion{Cl}{2+}/\ion{O}{+}) 
that can be applicable for {\hii} region conditions \citep{delgadoingladaetal15}. In Table~\ref{compicfcl} we compare our direct determinations of the Cl/H ratios for the  
{\hii} regions sample with the values we obtain using our quoted \ion{Cl}{2+}/\ion{H}{+} ratios and the $ICF$ schemes mentioned above. In Table~\ref{compicfcl} we also include 
the O/\ion{O}{2+} ratio as indicator of the ionization degree of the objects. In the case of the determinations based on the $ICF$ scheme by \citet{peimberttorrespeimbert77}, we have only included 
objects with $ICF$(\ion{Cl}{2+}) $<$ 2.0. In the case of the abundances obtained using the $ICF$ of \citet{mathisrosa91}, we use the coefficients for ``cool" Kurucz atmospheres for all the objects 
except for NGC~2579 and NGC~3603, for which ``hot" Kurucz atmospheres are more appropriate. The inspection of the values included in Table~\ref{compicfcl}, indicates that all the 
$ICF$ schemes provide Cl/H ratios systematically larger than our total Cl abundance determinations. All the schemes overestimate the contribution of \ion{Cl}{+} at lower 
ionization degrees but also of \ion{Cl}{3+} at higher ones. The correction by \citet{peimberttorrespeimbert77}, even of a very restricted applicability, gives values between 0.1 and 0.2 dex higher 
than our direct determinations. This is the $ICF$(\ion{Cl}{2+}) that gives the larger discrepancies for the low ionization objects. The rest of the $ICF$ schemes give differences smaller than 0.1 dex for the 
low ionization nebulae, those having O/\ion{O}{2+} $>$ 2. The situation becomes worst in the case of the high ionization {\hii} regions, the objects with O/\ion{O}{2+} $<$ 2. In this regime, the $ICF$ of \citet{mathisrosa91} 
give differences larger than 0.2 dex. The scheme of 
 \citet{izotovetal06} gives the best results, with differences smaller that 0.2 dex -- even smaller than 0.1 dex for most of the objects -- except in the case of NGC~3603, the object with the highest ionization. 
 The $ICF$ of \cite{delgadoingladaetal14} also gives reasonably good results with differences of the order or lower than 0.2 dex in all the cases. It is clear that all the $ICF$ schemes overestimate the 
 contribution of \ion{Cl}{3+} and that this difference increases with the ionization degree of the nebulae. 
 
Our results lead us to propose a new empirical $ICF$(\ion{Cl}{2+}) based on observations and valid for the ionization range covered by our objects. In Figure~\ref{fig_icf} we present the value of the 
$ICF$(\ion{Cl}{2+}) we obtain for each object -- which is simply the ratio between the total Cl abundance with respect to the \ion{Cl}{2+} abundance -- versus the ionization degree of the nebulae, 
parametrized by the O/\ion{O}{2+} ratio. The figure indicates a rather clear correlation that suggests nothing more complicated than a simple linear relation. The least-squares linear fitting is:
\begin{equation}
ICF(\mathrm{Cl}^{2+}) = 1.02 + 0.04 \times \mathrm{O}/\mathrm{O}^{2+}. 
\end{equation}
The uncertainty associated with our $ICF$(\ion{Cl}{2+}) is about 0.03 dex. The range of validity of this $ICF$(\ion{Cl}{2+}) correspond to the interval of O/\ion{O}{2+} ratio covered by our objects: 
1 $<$ O/\ion{O}{2+} $<$ 7.

%%%%%%%%%%%%%%%%%%%%%%%%%%%%%%%%%%%%%%%%%%%%%%%%%%%%%%%%%%%%%%%%%%%%%%%%%%%%%%%%%%%%%%%%%%%%%%%%%
\section{Conclusions} \label{conclusions}
%%%%%%%%%%%%%%%%%%%%%%%%%%%%%%%%%%%%%%%%%%%%%%%%%%%%%%%%%%%%%%%%%%%%%%%%%%%%%%%%%%%%%%%%%%%%%%%%%
 We determine the radial abundance gradient of Cl of the Milky Way from published spectra of {\hii} regions. The dataset is the deepest and most complete collection of spectra of Galactic {\hii} regions available. We have re-calculated the physical conditions -- electron temperature and density -- and ionic and total abundances of Cl and O using the same methodology and updated atomic data for all the objects. The total abundances of Cl and O are determined by simply adding 
their ionic abundances, without assuming ionization correction factors. 

We find that the slopes of the radial gradients of Cl and O are identical within the uncertainties: $-$0.043 dex kpc$^{-1}$. This is consistent with a lockstep evolution of both elements. There are previous determinations of the Galactic gradient of Cl but from observations of PNe. However, this is the first time that the gradient is computed from direct -- not $ICF$ based -- abundance determinations of Cl in {\hii} regions. We obtain that the mean value of the Cl/O ratio across the Galactic disc is log(Cl/O) = $-$3.42 $\pm$ 0.06, 
consistent with previous determinations based on PNe data \citep[e.g.][]{henryetal04} and from the recommended solar photospheric and meteoritic abundances compiled by \citet{loddersetal09}. 

We compare our direct determinations of the Cl/H ratios with those determined from our \ion{Cl}{2+} abundances and using some available $ICF$ schemes of the literature. We find that our total Cl abundances are always lower than the values determined using $ICF$s, indicating that those corrections schemes systematically overestimate the contribution of \ion{Cl}{+} and \ion{Cl}{3+} species to the total Cl abundance. Finally, we propose an empirical $ICF$(\ion{Cl}{2+}) to estimate the total Cl abundance 
in {\hii} regions when only the \ion{Cl}{2+}/ \ion{H}{+} ratio can be determined from their spectra. 

%%%%%%%%%%%%%%%%%%%%%%%%%%%%%%%%%%%%%%%%%%%%%%%%%%%%%%%%%%%%%%%%%%%%%%%%%%%%%%%%%%%%%%%%%%%%%%%%%
\section*{Acknowledgments}
We are grateful to the referee for his/her constructive comments. This paper is based on data obtained at the European Southern Observatory, Chile, with proposals ESO 68.C-0149(A), ESO 70.C-0008(A) and ESO 382.C-0195(A). This work has been funded by the Spanish Ministerio de Econom\'\i a y Competitividad (MINECO) under project AYA2011-22614. JGR acknowledges support from Severo 
Ochoa excellence program (SEV-2011-0187) postdoctoral fellowship.

%%%%%%%%%%%%%%%%%%%%%%%%%%%%%%%%%%%%%%%%%%%%%%%%%%%%%%%%%%%%%%%%%%%%%%%%%%%%%%%%%%%%%%%%%%%%%%%%%
%\bibliographystyle{mn2e}
%\bibliography{mnrasmnemonic,cesar_bibliography}

\label{lastpage}

\end{document}